

\magnification\magstep1
\TagsOnRight
\baselineskip=12pt
\NoBlackBoxes
\font\ninepoint=cmr9

\define\R{{\Bbb R}}            
meros reales
meros enteros

\def\k{\kappa}                 

\define\g{{\frak g}}           

do
do
do

do

            %
            %
           %
          %

\redefine\Ck{\text{\ \!C}}      
\redefine\Sk{\text{\ \!S}}      %
\define\>#1{{\bold#1}}                 
n para vectores

\define\co{\Delta}                     
\define\conm#1#2{\left[ #1,#2 \right]} 
\def\1{\'{\i}}                         

\font\titulo=cmbx10 scaled\magstep2
\font\cabeza=cmbx12


\def\Dr{1}
\def\Ji{2}
\def\Ti{3}
\def\Ruegg{4}
\def\TVi{5}
\def\TVii{6}
\def\TViii{7}
\def\Dob{8}
\def\IW{9}
\def\WB{10}
\def\BHOSi{11}
\def\BHOSii{12}
\def\BHOSiii{13}
\def\SHO{14}
\def\Ros{15}
\def\Nomi{16}
\def\Mas{17}
\def\Luk{18}


\
\vskip 3truecm

\centerline {\titulo  QUANTUM ALGEBRAS FOR MAXIMAL}
\bigskip
\centerline {\titulo  MOTION GROUPS OF N--DIMENSIONAL}
\bigskip
\centerline {\titulo  FLAT SPACES}
\bigskip

\vskip 1truecm

\centerline {A. Ballesteros, F.J. Herranz, M.A. del Olmo and M.
Santander} \bigskip\bigskip
\centerline{\it Departamento de F\1sica Te\'orica, Universidad de
                Valladolid.}
\centerline{\it E--47011 Valladolid, Spain.}
\smallskip
\centerline{e-mail: fteorica\@cpd.uva.es}

\vskip 2.5truecm

{\ninepoint
\noindent {\bf Abstract.}
An embedding method to get $q$-deformations for the non--semisimple
algebras generating the motion groups of $N$--dimensional flat spaces
is presented. This method gives a global and simultaneous scheme of
$q$-deformation for all $iso(p,q)$ algebras and for those ones obtained
from them by some In\"on\"u--Wigner contractions, such as the
$N$--dimensional Euclidean, Poincar\'e and Galilei algebras.

\vskip 1truecm \centerline {PACS: 02.10.T; 02.20.Sv; 03.65.Fd}
\bigskip
\medskip

\centerline {October 22,  1993.}

\bigskip

\centerline {Submitted to Lett. Math. Phys.}
\medskip

\vfill\eject

\noindent
\cabeza 1. Introduction\rm
\medskip
\medskip

The study of Drinfel'd--Jimbo \cite{\Dr,\Ji} $q$-deformations of
semisimple  Lie algebras has inspired the search for quantum algebras
corresponding to non--semisimple groups. Significant   but somehow
isolated results have been obtained in this field (cfr.
\cite{\Ti,\Ruegg} and references therein). In spite of the progress
recently done in the context of precising mathematically the
``universality" of these formal power series deformations
\cite{\TVi--\Dob} some kind of general framework encompassing these
quantum algebras  is  not available yet.  \medskip

We propose to base such a  scheme --for a large class of groups-- in
the classical structure underlying the geometries of real spaces with
projective metric (real Cayley--Klein geometries). For arbitrary
dimension, the $N$--dimensional ($N$D) Cayley--Klein (CK) groups are
the groups of motions of symmetric homogeneous $N$D real spaces with
maximal motion groups of dimension $\tfrac 12  N(N+1)$; this family
includes all the (pseudo)orthogonal groups $SO(p,q)$ $(p+q=N+1)$ (the
groups in Cartan series $B_l$ and $D_l$), as well as many other groups
obtained from them by a sequence of In\"on\"u--Wigner (IW) contractions
\cite{\IW}. Even among the groups got from the (pseudo)orthogonal ones
by a \it single \rm contraction, we find several interesting families
of groups, including the $ISO(p,q)$ $(p+q=N)$ and the groups
$T(SO(p,q)\otimes SO(p',q'))$  \cite{\WB}. In general, kinematical
groups also appear as Cayley--Klein groups.  \medskip

Starting from the case $N=2$, whose $q$-deformations are
well known, one could try to extend the quantum structure to higher
dimensions.  We have  completed   this programme   for the   $N=2,3$
cases \cite{\BHOSi,\BHOSii}, and the {\sl affine} $N=4$ case
\cite{\BHOSiii}. In this letter we report the  final result in the
quantization process as far as affine CK algebras are concerned. The
``naive" embedding of $(N-1)$D affine classical subalgebras into the
$N$D affine algebras for any $N\ge 4$ directly leads, through the
method described in Section 3, to a general $N$D Hopf algebra structure
(Section 4). The essentials on CK algebras needed to support this
approach are briefly summarized in Section 2, and some interesting
quantum algebras (Euclidean, Poincar\'e and Galilei ones, among others)
are discussed as particular cases in the concluding paragraph.

\bigskip
\bigskip

\noindent
\cabeza 2. Classical affine Cayley--Klein algebras\rm

\medskip\medskip

The $N$--dimensional CK algebras are real Lie algebras with dimension
$\tfrac 12 N(N+1)$, characterized by $N$ fundamental real parameters
$(\k_1,\dots,\k_N)$ \cite{\SHO}. We will denote them by
$\g_{(\k_1,\dots,\k_N)}$ and their associated  CK groups as

$G_{(\k_1,\dots,\k_N)}$. When all $\k_i$ are non zero we obtain the
$so(p,q)$ $(p+q=N+1,\, p\ge q\ge 0)$  algebras. When some $\k_i=0$ we
get the quasi--simple algebras obtained  from $so(p,q)$ through a
sequence of IW  contractions.  \medskip

Altogether, they are the Lie algebras of
the groups of motions of real spaces with a projective metric
\cite{\Ros}, and the link with the description that uses spaces is
established through a set of $N$ basic commuting involutions
$\Theta^{(M)}$ $(M=0,1,\dots,N-1)$ in $\g_{(\k_1,\dots,\k_N)}$. Each
involution $\Theta^{(M)}$ determines the Lie subalgebra $\frak
h^{(M)}\subset \g$ of all elements invariant under $\Theta^{(M)}$. The
dimension of $\frak h^{(M)}$ is  ${{M+1}\choose 2}+{{N-M}\choose 2}$.
The homogeneous spaces $\Cal X^{(M)}\equiv G/H^{(M)}$, where $H^{(M)}$
is the Lie subgroup of $G$ associated to $\frak h^{(M)}$, are symmetric
spaces with dimension $(M+1)(N-M)$, and $G$ acts transitively on each
of them. Moreover, they are identified with the spaces of points,
lines, 2-planes,\dots, $(N-1)$-planes if  $M=0,1,2,\dots,N-1$,
respectively.  \medskip In each space  $\Cal X^{(M)}$ there is a
canonical connection associated to the structure of symmetric space
\cite{\Nomi} and, in particular, the canonical conection in  $\Cal
X^{(0)}$ has constant curvature $\k_1$. By Cayley--Klein geometry one
should understand the whole set of spaces  $\Cal X^{(M)}$. \medskip

When $\k_1=0$, the space $\Cal X^{(0)}\equiv
G/H^{(0)}$ is flat, and each $N$--dimensional CK
algebra $\g_{(0,\k_2,\dots,\k_N)}$ can be realized as the algebra of
an {\sl affine} group of transformations on $\Bbb R^N$. Their
corresponding affine CK groups $G_{(0,\k_2,\dots,\k_N)}$ are generated
by $\langle P_i,J_{ij}\rangle$ $(i,j=1,\dots,N;\ i<j)$, and the Lie
brackets of $\g_{(0,\k_2,\dots,\k_N)}$ are given by $$
\aligned [P_i,P_j] &=0 ,\qquad
[J_{ij},P_k] =\delta_{ik}P_j-\delta_{jk}\k_{ij}\,P_i ,\\
[J_{ij},J_{lm}]&=\delta_{im}J_{lj}-\delta_{jl}J_{im}+
\delta_{jm}\k_{lm}J_{il}+\delta_{il}\k_{ij}J_{jm},\quad\ (i\le l,\
j\le m); \endaligned
\tag2.1
$$
where
$$
\k_{ij}=\prod_{s=i+1}^{j}\k_s,\quad i,j=1,\dots,N;\quad
(i<j).\tag2.2
$$
Each coefficient $\k_i$ can be scaled to $+1$, 0 or $-1$, so there are
$3^{N-1}$ different $N$D affine CK geometries.

{}From relations (2.1) it is clear that the groups
$G_{(0,\k_2,\dots,\k_N)}$ have a semidirect product structure:
$
G_{(0,\k_2,\dots,\k_N)}
\equiv T_N\odot G_{(\k_2,\dots,\k_N)},
$
where $T_N=\langle P_i \rangle$ $(i=1,\dots,N)$ is an Abelian subgroup
and $G_{(\k_2,\dots,\k_N)}=\langle J_{ij} \rangle$ $(i,j=1,\dots,N)$ is
a  $(N-1)$-dimensional CK group.
\medskip

One second order central element for the algebra
$\g_{(0,\k_2,\dots,\k_N)}$ coming from Killing's form reads
$$
\Cal C=P_N^2+\sum_{i=1}^{N-1}\k_{iN}P_i^2,\tag2.3
$$

A $(N+1)\times (N+1)$ matrix realization is given by means of the
matrices $e_{ij}$  with components
$(e_{ij})_{kl}=\delta_{ik}\delta_{jl}$ by
$$
P_i=e_{i0},\quad\ J_{ij}=-\k_{ij}e_{ij}+e_{ji},\qquad
i,j=1,\dots,N.\tag2.4
$$

A differential realization on $\Bbb L^2 (\Bbb R^N)$ associated with
the natural action of these groups on the homogeneous space
$G_{(0,\k_2,\dots,\k_N)}/G_{(\k_2,\dots,\k_N)}\simeq\R^N$ is:
$$
P_i=\partial_i,\quad\
J_{ij}=\k_{ij}x_j\partial_i  -  x_i\partial_j,\qquad i,j=1,\dots,N.
\tag2.5
$$
\medskip

The subalgebras  $\frak h^{(M)}$ of  $\g$ are:
$$
\frak h^{(M)}=\langle  P_i,J_{ij}\ (i,j=1,\dots,M);\ J_{lm}
(l,m=M+1,\dots,N)\rangle , \quad M=0,\dots,N-1.
\tag2.6
$$

The affine CK algebras are related by means of
IW contractions which will be denoted by $\Gamma^{(M)}$. The actions of
involution $\Theta^{(M)}$ and   contraction $\Gamma^{(M)}$ over a
generic element $X$ in the basis $\langle P_i,J_{ij}\rangle$ of $\g$
are:
$$
\align
\Theta^{(M)}(X)&=\left\{
\aligned
&X\quad \text {if}\ X\in \frak h^{(M)}\\
-&X\quad \text {if}\ X\notin \frak h^{(M)}
\endaligned
\right. ,\quad M=0, \dots,N-1;
\tag2.7\\
\Gamma^{(M)}(X)&=\left\{
\aligned
&X\quad \text {if}\ X\in \frak h^{(M)}\\
\varepsilon&X\quad \text {if}\ X\notin \frak h^{(M)}
\endaligned
\right. ,\quad M=1,\dots,N-1;\tag2.8
\endalign
$$
and extend by linearity to the whole algebra. The involutions generate
a $\Bbb Z_2^{\otimes N}$  abelian group, giving rise to a grading of
$\g_{(0,\k_2,\dots,\k_N)}$.  The contractions are obtained by making
the limit   $\varepsilon\to 0$ in (2.8); thus, we get a new CK algebra
with the parameter $\k_{M+1}=0$ whose associated  affine CK geometry
describes the behaviour of the original one in the neighbourhood of a
line, 2-plane,\dots, or $(N-1)$-plane, respectively.

\bigskip
\bigskip

\noindent
\cabeza 3. The deformation embedding method\rm

\medskip\medskip

In the classical construction, the embedding of $(N-1)$D affine CK
geometries within the $N$D ones can be used to derive many properties
of these geometrical systems. Our claim is that this idea gives rise
to  a general quantum structure for the affine CK algebras of arbitrary
dimension. The classical CK embedding can be stated as follows.

\proclaim{Proposition 3.1} The affine $N$--dimensional algebra
$\g_{(0,\k_2,\dots,\k_N)}$ contains $N$ affine subalgebras
$\Pi$ associated to $(N-1)$--dimensional  affine CK subgeometries and
generated by
$$ \Pi_{l_1 l_2\dots l_{N-1}}=
\langle P_i,J_{ij}\rangle,\quad l_s\in\{1,\dots,N\}, \quad l_s<l_{s+1}
\quad i,j=l_1,\dots,l_{N-1},\tag3.1
$$
where each possible set $l_1 l_2\dots l_{N-1}$ is obtained by taking
out a single index $r$ in the sequence $1,2,\dots,N$.

\endproclaim

The generators of the  subalgebra $\Pi_{l_1 l_2\dots l_{N-1}}$ are
obtained by removing from those of $g_{(0,\k_2,\dots,\k_N)}$ the
translation generator $P_r$ as well as all rotations $J_{ij}$ which
have either $i=r$ or $j=r$. The algebra closed by the remaining
generators has still  the form (2.1) provided the set of fundamental
parameters associated to $\Pi_{l_1 l_2\dots l_{N-1}}$ is taken as
$(0,\k_{l_1l_2},\dots,\k_{l_{N-2}l_{N-1}})$. If $r=N$ we get directly
the  $(N-1)$D affine CK  algebra with fundamental constants
$(0,\k_{12},\dots,\k_{{(N-2)}{(N-1)}})\equiv(0,\k_2,\dots,\k_{N-1})$.
On the other hand, if $r<N$  the last parameter  $\k_N$ always appears
in the Lie brackets of  $\Pi_{l_1 l_2\dots l_{N-1}}$.

\medskip

Let us take the $N$D affine CK algebra $\g_{(0,\k_2,\dots,\k_N)}$
together with the quantum deformation of the $(N-1)$D case $U_q
\g_{(0,\k_2,\dots,\k_{N-1})}$.  Consider the $N-1$ affine subalgebras
$\Pi_{l_1,\dots,l_{N-1}}$ which include $P_N$ (and hence $\k_N$) and
write the coproducts and deformed commutation relations for these
subalgebras as $(N-1)$D quantum universal enveloping CK algebras
$U_q(\Pi_{l_1,\dots,l_{N-1}}) \equiv U_q
\g_{(0,\k_{l_1l_2},\dots,\k_{l_{N-2}l_{N-1}})}$. Then,

\roster

\item Assume that $U_q(\Pi_{l_1,\dots,l_{N-1}})$ are restrictions of
the quantization for $U_q \g_{(0,\k_2,\dots,\k_{N})}$. We understand by
restriction the quantum universal enveloping algebra analogue of the
geometrical projection along $P_r$: the generators of $U_q
\g_{(0,\k_2,\dots,\k_{N})}$ not included in  $\Pi_{l_1,\dots,l_{N-1}}$
are necessarily taken as zero.

\item Take the coproduct and algebra relations of $U_q
\g_{(0,\k_2,\dots,\k_{N})}$ as the simplest ones consistent with these
restrictions.

\item Check the selfconsistency of the resultant structure by imposing
that the coproduct is a Hopf algebra homomorphism.

\endroster

This method has been used in \cite{\BHOSiii} to derive the $N=4$
quantum affine  algebras in terms of the known $N=3$ ones
\cite{\BHOSii}. In that case, the $N=4$  coproduct is completely
determined by the comultiplications of the $N=3$ restrictions. However,
some additional terms must be added in the commutation relations coming
from the subalgebras in order to fulfill step (3). It is also possible
to view in the same light the construction \cite{\BHOSi} of the 3D
affine case starting from the 2D one. Now, the homomorphism constrain
also imposes some new additional terms in the 3D coproducts which have
no 2D counterparts.

\medskip

This paper shows  that the situation is completely different for
dimensions higher than 4. Starting from the $N=4$ case and using an
induction method, the procedure outlined above does define both the
coproduct for the generic $N$D   case as well as the corresponding
deformed brackets. No additional terms have to be included to ensure
the homomorphism property of the Hopf structure. Thus, the 4D case is
the cornerstone of the Hopf structure for this family of algebras.
\medskip

We give a sketch of the method for getting the 5D case
that illustrates the main features of the $N$D expressions that will
be explicitly given in the next section.

\medskip

Recall that the $N=4$ quantum affine algebras was defined by the
coproduct \cite{\BHOSiii}
$$
\aligned
&\qquad\qquad\qquad\co(X)=1\otimes X+ X\otimes 1,\quad
X\in\{P_4;J_{12},J_{13},J_{23}\},\\
&\qquad\qquad\qquad\co(P_i)=e^{-\frac z2 P_4}\otimes P_i +P_i\otimes
e^{\frac z2 P_4},  \quad i=1,2,3;\\ \co(J_{14})&=e^{-\tfrac z2
P_4}\otimes J_{14} +J_{14}\otimes e^{\tfrac z2 P_4} +\tfrac z2
J_{12}e^{-\tfrac z2 P_4}\otimes \k_3\k_4 P_2 - \k_3\k_4   P_2 \otimes

e^{\tfrac z2 P_4} \tfrac z2 J_{12}\\ &\qquad\qquad\qquad\qquad
+\tfrac z2 J_{13}e^{-\tfrac z2 P_4}\otimes \k_4 P_3
- \k_4   P_3 \otimes   e^{\tfrac z2 P_4} \tfrac z2 J_{13},\\
\co(J_{24})&=e^{-\tfrac z2 P_4}\otimes J_{24} +J_{24}\otimes e^{\tfrac
z2 P_4} -\tfrac z2 J_{12}e^{-\tfrac z2 P_4}\otimes \k_3\k_4 P_1
+\k_3\k_4  P_1 \otimes   e^{\tfrac z2 P_4}\tfrac z2  J_{12}\\
&\qquad\qquad\qquad\qquad
+\tfrac z2 J_{23}e^{-\tfrac z2 P_4}\otimes \k_4 P_3
- \k_4 P_3 \otimes   e^{\tfrac z2 P_4} \tfrac z2   J_{23},\\
\co(J_{34})&=e^{-\tfrac z2 P_4}\otimes J_{34} +J_{34}\otimes e^{\tfrac
z2 P_4} -\tfrac z2 J_{13}e^{-\tfrac z2 P_4}\otimes \k_4 P_1
+\k_4  P_1 \otimes   e^{\tfrac z2 P_4}\tfrac z2  J_{13}\\
&\qquad\qquad\qquad\qquad
-\tfrac z2 J_{23}e^{-\tfrac z2 P_4}\otimes\k_4 P_2
+ \k_4  P_2 \otimes   e^{\tfrac z2 P_4} \tfrac z2 J_{23},
\endaligned\tag3.2
$$
and the following deformed brackets
$$
[J_{14},P_{1}]=\Sk_{-z^2}(P_4),\quad
[J_{24},P_{2}]=\Sk_{-z^2}(P_4),\quad
[J_{34},P_{3}]=\Sk_{-z^2}(P_4),\tag3.3a
$$
$$
\aligned
[J_{14},J_{24}]&=\k_3\k_4\left\{J_{12}\Ck_{-z^2}(P_4)
+\tfrac{z^2}4\k_4P_3W_{123}\right\},\\
 [J_{14},J_{34}]&=\k_4\left\{J_{13}\Ck_{-z^2}(P_4)
-\tfrac{z^2}4\k_3\k_4P_2W_{123}\right\},\\

[J_{24},J_{34}]&=\k_4\left\{J_{23}\Ck_{-z^2}(P_4)
+\tfrac{z^2}4\k_3\k_4P_1W_{123}\right\}.
\endaligned \tag3.3b
$$
where $W_{123}=\k_2P_1J_{23}-P_2J_{13}+P_3 J_{12}$ is a second order
Casimir of $\Pi_{123}$, and
$$
\Ck_{-z^2}(X)=\frac{e^{z X}+e^{-z X}}{2}, \qquad
\Sk_{-z^2}(X)=\frac{e^{z X}-e^{-z X}}{2z} \tag 3.4
$$
are used for the sake of consistency with the general CK quantum scheme
\cite{\BHOSi,\BHOSii}.
\medskip

The $N=5$ affine CK  geometries have five affine CK
4D subgeometries that are generated by the following generators and
characterized by their associated set of fundamental $\kappa_i$
constants   $$
\alignedat2
\Pi_{1234}&  =\langle
P_1,P_2,P_3,P_4,J_{12},J_{13},J_{14},J_{23},J_{24},J_{34}\rangle,
&\qquad &(0,\k_2,\k_3,\k_4);\\
\Pi_{1235}&  =\langle
P_1,P_2,P_3,P_5,J_{12},J_{13},J_{15},J_{23},J_{25},J_{35}\rangle,
&\qquad &(0,\k_2,\k_3,\k_4\k_5);\\
\Pi_{1245}&  =\langle
P_1,P_2,P_4,P_5,J_{12},J_{14},J_{15},J_{24},J_{25},J_{45}\rangle,
&\qquad &(0,\k_2,\k_3\k_4,\k_5);\\
\Pi_{1345}&  =\langle
P_1,P_3,P_4,P_5,J_{13},J_{14},J_{15},J_{34},J_{35},J_{45}\rangle,
&\qquad &(0,\k_2\k_3,\k_4,\k_5);\\
\Pi_{2345}&  =\langle
P_2,P_3,P_4,P_5,J_{23},J_{24},J_{25},J_{34},J_{35},J_{45}\rangle,
&\qquad &(0,\k_3,\k_4,\k_5).
\endalignedat
\tag3.5
$$
All these 4D subalgebras can be immediately quantized by means of
(3.2) and (3.3): we have only to rename generators following the
pattern given by $\Pi_{1234}$ and enter the appropriate $\k_i$
coefficients.  \medskip

Now, we consider the quantization of the
four subalgebras involving $\k_5$ as restrictions of the general
structure of the 5D case. For instance, $J_{25}$ belongs to
$\Pi_{1235},\Pi_{1245}$ and $\Pi_{2345}$. Hence, we can write for it
three associated coproducts as different restrictions (in the 5D
quantum algebra) of $\co(J_{25})$ :  $$
\aligned
\co(J_{25})\big|_{\Pi_{1235}}\!&=e^{-\tfrac z2 P_5}\!\otimes\! J_{25}
+J_{25}\!\otimes \!e^{\tfrac z2 P_5} -\tfrac z2 J_{12}e^{-\tfrac z2
P_5}\!\otimes\! \k_3\k_4\k_5 P_1 +\k_3\k_4\k_5  P_1 \!\otimes \!
e^{\tfrac z2 P_5}\tfrac z2  J_{12}\\ &\qquad\qquad\qquad\qquad
+\tfrac z2 J_{23}e^{-\tfrac z2 P_5}\otimes \k_4\k_5 P_3
- \k_4\k_5 P_3 \otimes   e^{\tfrac z2 P_5} \tfrac z2   J_{23},\\

\co(J_{25})\big|_{\Pi_{1245}}\!&=e^{-\tfrac z2 P_5}\!\otimes \!J_{25}
+J_{25}\!\otimes \!e^{\tfrac z2 P_5} -\tfrac z2 J_{12}e^{-\tfrac z2
P_5}\!\otimes\! \k_3\k_4\k_5 P_1 +\k_3\k_4\k_5  P_1 \!\otimes \!
e^{\tfrac z2 P_5}\tfrac z2  J_{12}\\ &\qquad\qquad\qquad\qquad
+\tfrac z2 J_{24}e^{-\tfrac z2 P_5}\otimes \k_5 P_4
- \k_5 P_4 \otimes   e^{\tfrac z2 P_5} \tfrac z2   J_{24},\\
\co(J_{25})\big|_{\Pi_{2345}}&=e^{-\tfrac z2 P_5}\otimes J_{25}
+J_{25}\otimes e^{\tfrac z2 P_5} +\tfrac z2 J_{23}e^{-\tfrac z2
P_5}\otimes \k_4\k_5 P_3 - \k_4\k_5   P_3 \otimes   e^{\tfrac z2 P_5}
\tfrac z2 J_{23}\\ &\qquad\qquad\qquad\qquad
+\tfrac z2 J_{24}e^{-\tfrac z2 P_5}\otimes \k_5 P_4
- \k_5   P_4 \otimes   e^{\tfrac z2 P_5} \tfrac z2 J_{24}.
\endaligned
\tag3.6
$$
We choose for the complete coproduct for $J_{25}$ the simplest one
which is consistent with these restrictions:
$$
\aligned
\co(J_{25})&=e^{-\tfrac z2 P_5}\otimes J_{25} +J_{25}\otimes
e^{\tfrac z2 P_5} -\tfrac z2 J_{12}e^{-\tfrac z2 P_5}\otimes
\k_3\k_4\k_5 P_1 +\k_3\k_4\k_5  P_1 \otimes   e^{\tfrac z2 P_5}\tfrac
z2  J_{12}\\ &\qquad\qquad\qquad\qquad
+\tfrac z2 J_{23}e^{-\tfrac z2 P_5}\otimes \k_4\k_5 P_3
- \k_4\k_5 P_3 \otimes   e^{\tfrac z2 P_5} \tfrac z2   J_{23},\\

&\qquad\qquad\qquad\qquad
+\tfrac z2 J_{24}e^{-\tfrac z2 P_5}\otimes \k_5 P_4
- \k_5   P_4 \otimes   e^{\tfrac z2 P_5} \tfrac z2 J_{24}.
\endaligned
\tag3.7
$$

As far as the deformed brackets are concerned, we examine --for
instance-- the case of $[J_{15},J_{25}]$. This bracket can be found
both in $\Pi_{1235}$ and in $\Pi_{1245}$. The respective restrictions
read  $$
\aligned
[J_{15},J_{25}]\big|_{\Pi_{1235}}
&=\k_3\k_4\k_5\left\{J_{12}\Ck_{-z^2}(P_5) +\tfrac{z^2}4\k_4\k_5
P_3W_{123}\right\},\\ [J_{15},J_{25}]\big|_{\Pi_{1245}}
&=\k_3\k_4\k_5\left\{J_{12}\Ck_{-z^2}(P_5) +\tfrac{z^2}4 \k_5
P_4W_{124}\right\}, \endaligned
\tag3.8
$$
where $W_{124}=\k_2P_1J_{24}-P_2J_{14}+P_4 J_{12}$. The Ansatz (2)
leads to the following deformed commutation rule in $U_q
\g_{(0,\k_2,\dots,\k_{N})}$: $$
[J_{15},J_{25}]=\k_3\k_4\k_5\left\{J_{12}\Ck_{-z^2}(P_5)
+\tfrac{z^2}4\left(\k_4\k_5 P_3W_{123}+\k_5 P_4W_{124}\right)
\right\}. \tag3.9
$$

The rather remarkable result we get is  that the repeated use of this
procedure gives rise to a coproduct and commutation relations which
automatically makes the comultiplication a Hopf homomorphism, so no new
terms appear either in the coproduct or in the Lie brackets when going
from $N=4$ to $N=5$; the full set of expressions obtained as (3.7) and
(3.9) can be easily proved to define by themselves a right quantization
for $N=5$.

\bigskip
\bigskip

\noindent
\cabeza 4. Quantum affine Cayley--Klein algebras\rm

\medskip\medskip

It is only matter of a straightforward computation and induction to
show that the same result  holds when going from $N$ to $N+1, N\ge
5$, following the same pattern described in \S 3. The final
expresions for $U_q\frak
g_{(0,\k_2,\dots,\k_N)}$, which include as well the $N=2,3,4$ cases,
are given in the:

\noindent
\proclaim{Theorem 4.1}
The quantized universal enveloping affine CK algebra
$U_q\frak g_{(0,\k_2,\dots,\k_N)}$ ($N\ge 2$) is defined by:
\medskip

\noindent
(1) Coproduct:
$$
\aligned
\co(X)&=1\otimes X+ X\otimes 1,
\qquad X\in\{P_N;J_{ij},\, i,j=1,\dots,N-1\},\\
\co(P_i)&=e^{-\frac z2 P_N}\otimes P_i +
P_i\otimes e^{\frac z2 P_N},
\qquad i=1,\dots,N-1;\\
\co(J_{iN})&=e^{-\tfrac z2 P_N}\otimes J_{iN} +J_{iN}\otimes e^{\tfrac
z2 P_N}\\ &\quad
-\sum_{s=1}^{i-1}\tfrac z2 J_{si}e^{-\tfrac z2 P_N}\otimes \k_{iN} P_s
+\sum_{s=1}^{i-1}\k_{iN} P_s \otimes e^{\tfrac z2 P_N} \tfrac z2
J_{si},\\ &\quad
+\sum_{s=i+1}^{N-1}\tfrac z2 J_{is}e^{-\tfrac z2 P_N}\otimes \k_{sN}
P_s -\sum_{s=i+1}^{N-1}\k_{sN} P_s \otimes e^{\tfrac z2 P_N} \tfrac z2
J_{is}, \quad i=1,\dots,N-1.
\endaligned\tag4.1
$$

\noindent
(2)  Counit:
$\epsilon(P_i)=\epsilon(J_{ij})=0,\quad i,j=1,\dots,N.$\hfill $(4.2)$

\noindent
(3) Antipode:
$$
\aligned
\gamma(X)&=-e^{(N-1)\tfrac {z}2 P_N}X e^{-(N-1)\tfrac {z}2 P_N}\\
&=\cases -P_i,\qquad  i=1,\dots,N,\qquad\qquad {\text{if}}\ X=P_i,\\
-J_{lm},\qquad  l,m=1,\dots,N-1,\qquad\quad {\text{if}}\ X=J_{lm} ,\\
-J_{lN}-\k_{lN}(N-1)\tfrac {z}2P_l,\quad
 l=1,\dots,N-1,\quad {\text{if}}\ X=J_{lN} .\endcases
\endaligned
\tag 4.3
$$

\comment
explicitly,
$$
\aligned
\gamma(P_i)&=-P_i,\qquad  i=1,\dots,N;\\
\gamma(J_{lm})&=-J_{lm},\quad
\gamma(J_{lN})=-J_{lN}-\k_{lN}(N-1)\tfrac {z}2P_l,\quad
 l,m=1,\dots,N-1.
\endaligned
\tag4.3b
$$
\endcomment

\noindent
(4) The deformed algebra relations are (the remaining ones are
undeformed and given by (2.1)):
$$
\aligned
[J_{iN},P_{i}]&=\Sk_{-z^2}(P_N),\quad i=1,\dots,N-1,\\
[J_{iN},J_{jN}]&=\k_{jN}\biggl\{J_{ij}\Ck_{-z^2}(P_N)\\
&+\dfrac{z^2}4\biggl(\sum_{s=1}^{i-1}\k_{iN}P_sW_{sij}
-\sum_{s=i+1}^{j-1}\k_{sN}P_sW_{isj}+\sum_{s=j+1}^{N-1}
\k_{sN}P_sW_{ijs}\biggr)    \biggr\}\quad i<j, \endaligned \tag4.4 $$
where $$ W_{ijk}=\k_{ij}P_iJ_{jk}-P_jJ_{ik}+P_kJ_{ij},\quad i<j<k,\quad
i,j,k=1,\dots,N-1. \tag4.5
$$
\endroster
\endproclaim

\noindent
{\it Remarks:}
\item {i)} The three--index symbols $W_{ijk}$ are
the second quadratic Casimir operator in the classical 3D subalgebra
$\Pi_{ijk}$ (the first quadratic central element is given by (2.3));
and it appears as the Pauli--Lubanski operator in the (2+1) Poincar\'e
group (hence the name of $W$).

\item{ii)} The $J_{ij}$ $(i,j=1,\dots,N-1)$ close a
$(N-2)$D {\sl classical} CK algebra, with fundamental coefficients
$(\k_2, \dots, \k_N)$. The translations commute, and only $P_N$ is
primitive.

\item{iii)} An implicit rule to be followed when applying expressions
(4.1--5) to each specific case, in accordance with current usage, is
that sums whose range is empty are equal to zero. The terms including
$W_{ijk}$ in (4.4) do not appear in the cases $N=2,3$ as it is implied
by (4.5) and by the fact that $W_{ijk}$ requires three different
indices.

\proclaim{Proposition 4.2} The quantum analogue of the second  order
Casimir (2.3) is $$
\Cal C^q=4\left[\Sk_{-z^2}(\tfrac 12
P_N)\right]^2+\sum_{i=1}^{N-1}\k_{iN}P_i^2. \tag4.6
$$
\endproclaim

\proclaim{Proposition 4.3} A differential realization of (4.4) is
given by $$
\alignedat2
P_i&=\partial_{i},&\quad i&=1,2,\dots,N;\\
J_{lm}&=\k_{lm} x_m\partial_{l}  - x_l\partial_{m} ,&\quad
l,m&=1,2,\dots, N-1;\\
J_{lN}&=\k_{lN}x_N  \partial_{l} - x_l\Sk_{-z^2}(\partial_{N})
,&\quad l&=1,2,\dots, N-1.
\endalignedat
\tag4.7
$$
\endproclaim
Note that, for this realization all $W_{ijk}\equiv 0$.
\medskip

The proofs of all these statements together with the fact that the
classical limit $z\to 0$ leads to the expressions  given in \S 2 is a
matter of computation.

\medskip

Involutions and IW  contractions associated to the algebraic
structure of the classical CK algebras can be implemented  by
introducing a transformation of the deformation parameter $z$. So, a
$q$-involution $\Theta_q^{(M)}$ and a $q$-contraction $\Gamma_q^{(M)}$
are defined by
$$
\align
\Theta_q^{(M)}(X,z)&:=(\Theta^{(M)}(X),-z),\ \ M=0,\dots,N-1;\tag4.8\\
\Gamma_q^{(M)}(X,z)&:=(\Gamma^{(M)}(X),z/\varepsilon),\ \
M=1,\dots,N-1; \tag4.9
\endalign
$$
where
$\Theta^{(M)}$ is the classical involution  (2.7) and $\Gamma^{(M)}$
the  IW contraction (2.8). The $q$-involutions (4.8) generate again an
Abelian group $(\Bbb Z_2^{\otimes N})$ which leaves invariant the Hopf
algebra (4.1--4.5). On the other hand, the transformation
$\Gamma_q^{(M)}$ corresponds to the limit $\k_{M+1}\rightarrow 0$.
\bigskip
\medskip

\noindent
\cabeza 5. Some examples\rm

\medskip\medskip

We would like to emphasize that the Hopf structure (4.1--4.5) allows
to get a $q$-deforma-tion for  a rather large set of non semisimple Lie
algebras.  In the following we discuss some particular  interesting CK
groups  whose associated quantum algebras are described by Theorem 4.1.

\medskip

\noindent
(1) When $\k_i\ne 0$ $(i=2,\dots,N)$ we get the inhomogenous
(pseudo)orthogonal  groups $ISO(p,q)$ with $p\ge q\ge 0$ and $p+q=N$.
These groups act on the space of points  by isometries of the metric
induced in $\Cal X^{(0)}\equiv G/H^{(0)}$ by the bilinear form
$\Lambda$ on $\Bbb R^N$: $$
\Lambda=\text{diag}\,(1,\k_{12},\k_{13},\dots,\k_{1N}) =
(1,\k_2,\k_2\k_3,\dots,\k_2 \cdots \k_{N}). \tag5.1
$$
Thus, we find for $N=5$ that the Euclidean group $(ISO(5))$ apears once
for the values of the coeficients $(0, \k_2, \k_3, \k_4, \k_5)$ equal
to $(0,+,+,+,+)$ (hence $\Lambda=\text{diag}\,(1,+,+,+,+)$) while the
(4+1) Poincar\'e group $(ISO(4,1))$ appears five times: $(0,-,+,+,+)$,
$(0,+,+,+,-)$, $(0,-,-,+,+)$, $(0,+,+,-,-)$ and $(0,+,-,-,+)$. In
particular, the quantum algebra labelled by $(0,+,\dots,+,-)$ is the
$N$--dimensional version of the $\k$--Poincar\'e given in [\Mas,\Luk].
\medskip

\noindent
(2) When some $\k_i=0$ there exist different kinds of groups. Their
action is always well defined although it is not realized  by
isometries of a non--degenerate metric in $\Bbb R^N$ \cite{\SHO}. It is
worth to mention three types of groups for a {\sl unique} coefficient
$\k_i$ zero:  \roster

\item"{(2.1)}" $\k_2=0$: it gives rise to twice inhomogeneous
(pseudo)orthogonal groups $IISO(p,q)$ with $p+q=N-1$; e.g. $IISO(4)$ is
the (4+1) Galilean group defined by $(0,0,+,+,+)$.
\smallskip

\item"{(2.2)}" $\k_N=0$: it originates once inhomogeneous and
once contragredient inhomogeneous (pseudo)orthogonal groups
$II'SO(p,q)$ with $p+q=N-1$; e.g. $II'SO(4)$ is the (4+1) Carroll group
determined by $(0,+,+,+,0)$. \smallskip

\item"{(2.3)}" $\k_i=0$ $(i\ne 2,N)$: it corresponds to  the
inhomogeneous $IT_r(SO(p,q)\otimes SO(p',q'))$ of the groups
$T_r(SO(p,q)\otimes SO(p',q'))$ studied in [\WB]. For $\k_3=0$ they are
$IT_{2(N-2)} (SO(p,q)\otimes SO(p',q'))$ with $p+q=N-2$ and $p'+q'=2$.
\endroster

\noindent
(3) The extreme case is when {\sl all} $\k_i=0$; this group
corresponds to the flag space $F_N$.
\medskip

Finally, we would like to point out some aspects of the results  we
have outlined above. First, they are found by means of a direct
constructive method:  $q$-deformations for any dimension are obtained
from the lower dimensional ones. Second,  an overall global view of all
affine $N$--dimensional algebras is provided without making any
recourse to complex forms or to special transformations, so it is
self--contained. Third,  new quantizations are obtained and most of the
already known ones are included (see \cite{\BHOSi--\BHOSiii} and
references therein).  \medskip

The theory  developed here for the affine case raises further hope
that this method could lead to an explicit Hopf structure for the
general case of Cayley--Klein groups associated to spaces of non--zero
constant curvature ($\k_1 \neq 0$). Work on this line is currently in
progress.

\bigskip
\bigskip

\noindent{\bf Acknowledgments}
\medskip

This work has been partially supported by a DGICYT project
(PB91--0196) from the Ministerio de Educaci\'on y Ciencia de
Espa\~na.

\bigskip
\bigskip

\noindent{\bf References}

\eightpoint
\medskip

\ref      
\no{\Dr}
\by  Drinfeld V. G.
\jour Proceedings of the International Congress of Mathematics,
MRSI Berkeley, (1986) 798
\endref

\ref      
\no{\Ji}
\by Jimbo M.
\jour Lett. Math. Phys.
\yr 1985
\vol 10
\pages 63

\moreref
\yr 1986
\vol 11
\pages 247
\endref

\ref     
\no{\Ti}
\by      Celeghini E.,  Giachetti R.,  Sorace E., and  Tarlini M.
\paper   Contractions of quantum groups
\jour    Lecture Notes in Mathematics n. 1510, 221,
         Springer-Verlag, Berl\1n (1992)
\endref

\ref     
\no{\Ruegg}
\by        Ruegg H.
\paper   $q$-Deformations of semisimple and non--semisimple Lie
algebras \jour    Integrable Systems, Quantum groups and Quantum Field
Theories
         NATO ASI Series
\yr       1993
\pages    45
\endref

\ref      
\no{\TVi}
\by  Truini P., and  Varadarajan V. S.
\jour Lett. Math. Phys.
\vol 24
\yr 1992
\pages 63
\endref

\ref      
\no{\TVii}
\by  Truini P., and  Varadarajan V. S.
\jour Lett. Math. Phys.
\yr 1992
\vol  26
\pages 53
\endref

\ref      
\no{\TViii}
\by  Truini P., and  Varadarajan V. S.
\jour Rev. Math. Phys.
\yr 1993
\vol 5
\pages 363
\endref

\ref        
\no{\Dob}
\by         Dobrev V. K.
\jour        Proceedings of the  XIX ICGTMP
\publ       Anales de F\1sica, Monograf\1as. Vol. 1.I, p. 91.
CIEMAT/RSEF \publaddr   Madrid (1993)
\yr
\endref

\ref       
\no{\IW}
\by        In\"on\"u E., and  Wigner E. P.
\jour      Proc. Natl. Acad. Sci. U. S.
\vol       39
\yr        1953
\pages     510
\endref

\ref      
\no{\WB}
\by         Wolf K .B., and  Boyer C. B.
\jour      J. Math. Phys.
\vol       15
\yr        1974
\pages     2096
\endref

\ref        
\no{\BHOSi}
\by          Ballesteros A.,  Herranz F. J.,  del Olmo M. A., and
Santander M. \jour       J. Phys. A
\vol        26
\yr         1993
\pages      5801
\endref

\ref        
\no{\BHOSii}
\by          Ballesteros A.,  Herranz F. J.,  del Olmo M. A., and
Santander M. \jour       J. Phys. A.
\vol         27
\yr         1994
\pages       1283
\endref

\ref        
\no{\BHOSiii}
\by          Ballesteros A.,  Herranz F. J.,  del Olmo M. A., and
Santander M. \paper      4D quantum affine algebras and space--time
$q$-symmetries \jour       preprint hep--th/9310140
\vol
\yr         1993
\endref

\ref        
\no{\SHO}
\by         Santander M.,  Herranz F. J., and  del Olmo M. A.

\moreref
\jour        Proceedings of the  XIX ICGTMP
\publ       Anales de
F\1sica, Monograf\1as. Vol. 1.I, p. 455.  CIEMAT/RSEF \publaddr
Madrid (1993) \yr
\endref

\ref      
\no{\Ros}
\by       Yaglom I. M.,  Rozenfel'd B. A., and Yasinskaya  E. U.
\jour      Sov. Math. Surveys
\vol         19
\vol         n5
\pages     49
\yr         1966
\endref

\ref      
\no{\Nomi}
\book     Foundations of differential geometry
\by        Kobayashi S., and  Nomizu K.
\yr       1963
\bookinfo Vol. 1, Wiley Interscience,  New York
\endref

\ref        
\no{\Mas}
\by         Maslanka, P.
\jour       J. Phys. A
\vol        26
\yr         1993
\pages      L1251
\endref

\ref        
\no{\Luk}
\by          Lukierski, J. and Ruegg, H.
\jour       University of Geneva preprint
\vol
\yr         1993
\endref

\end